\documentstyle[12pt]{article}
\global\arraycolsep=2pt 
 \newcommand{\be}{\begin{eqnarray}}
\newcommand{\ee}{\end{eqnarray}}
\newcommand{\la}{\langle}
\newcommand{\ra}{\rangle}
\begin{document}
\begin{titlepage}
 \vspace{0.3cm}
\begin{center}
\Large\bf  
How strange a non-strange  heavy baryon?
\end{center}
\vspace {0.3cm}

 \begin{center}    {\bf Ariel R. Zhitnitsky}\footnote{
  e-mail address:arz@physics.ubc.ca }
 \end{center}
\begin{center}
{\it Physics Department, University of British Columbia,
6224 Agricultural  Road, Vancouver, BC V6T 1Z1, Canada }\\  
and \\ 
{\it Budker Inst. of Nuclear Physics, Novosibirsk, 630090, Russia}
\end{center}
 \begin{abstract}
We give some general arguments in favor of the large magnitude of  
matrix elements of an operator associated with nonvalence quarks
in heavy hadrons.  
We estimate matrix element 
$\frac{1}{2m_{\Lambda_b}}\la \Lambda_b|\bar{s}s|\Lambda_b\ra 
 \simeq 1\div 2$ 
    for $\Lambda_b$ baryon
 whose valence content is $b, u, d$       quarks.  
This magnitude corresponds to a noticeable contribution of the strange quark
into the heavy baryon mass $\frac{1}{2m_{\Lambda_b}}\la \Lambda_b|m_s\bar{s}s|\Lambda_b\ra\simeq 200\div 300 MeV$.
The arguments are based on the QCD sum rules and low energy theorems.
The physical picture behind of the phenomenon is somewhat
 similar to the one associated with the   large strange
content of the nucleon where  matrix
element  $  \la p|\bar{s}s|p\ra\simeq 1$ by no means is small.
We discuss some possible  applications of the result.

\end{abstract}
\end{titlepage}

\section{Introduction and Motivation.}
Nowadays it is almost accepted that a {\bf nonvalence} component
in a hadron could be very high, much higher than naively
one could expect from the naive perturbative estimations.
Experementally, such a phenomenon was observed in a number
of places. Let me mention only few of them.

First of all it is anomalies in charm hadroproduction.
As is known, the cross section
for the production of $J/\psi 's$ at high transverse momentum
at the Tevatron is a factor $\sim 30$ above 
the standard perturbative QCD predictions. The production cross sections
for other heavy quarkonium states also show similar anomalies\cite{Abe}.

The second example of the same kind is the charm structure function
of the proton  measured by EMC collaboration \cite{Aubert} is some 
$30$ times larger at $x_{Bj}=0.47,~Q^2=75 GeV^2$ than that predicted 
on  the standard calculation of photon-gluon fusion $\gamma^{ast}g
\rightarrow c\bar{c}$.

Next example    is  the
matrix element $\la N|\bar{s}s|N\ra$ which
  does not vanish, as naively one could expect,     
 but rather, has the same order of magnitude as 
valence matrix element $\la N|\bar{d}d|N\ra$.

One can present many examples of such a kind, where ``intrinsic''
non-valence component plays an important role.
 This is not the place
to analyze all these unexpected deviations 
from the standard perturbative predictions. 
The only point we would like to make
here is the following. Few examples mentioned above ( for  
more examples see  recent review   \cite{Brodsky1})
unambiguously suggest that a non-valence component in a hadron 
in general is not small.
In QCD-terms it means that the corresponding matrix element
has non-perturbative origin 
and has no $\alpha_s$ suppression which is naively expected
from perturbative analysis (we use the term ``intrinsic  component''
to describe this non-perturbative contribution in order 
to distinguish from  the ``extrinsic component'' which is always present
and is nothing but a perturbative amplitude of the gluon splitting $g\rightarrow Q\bar{Q}$
 with  non-valence quark flavor $Q$  ). 
 
The phenomenon we are going to discuss here is somewhat similar 
to those effects mentioned above.   
We shall argue  that
  a non-valence component
in  a heavy-light quark system could be very large.   However, before
to present our argumentation  of  why, let say, the matrix element
$\la \Lambda_b|\bar{s}s|\Lambda_b\ra$  is not suppressed
(i.e. has the same order of magnitude as 
valence matrix element $\la \Lambda_b|\bar{u}u|\Lambda_b\ra$),
 we would like  to get some  QCD-based explanation  
of the similar effects  we   mentioned earlier.  

Before to do so, let me remind that
 for a long time it was widely believed that the admixture of the pairs of
non-valence quarks in hadrons
 is small. The main justification
of  this  picture was the constituent quark model where 
there is no room, let say, for a strange quark in the nucleon,
(see,  however, the recent paper \cite{Isgur} on this subject).
  It has been known for a while that  this picture is not quite true:
In scalar and pseudoscalar   channels
one can expect a noticeable deviation from this naive prediction.
This is because,
these channels are very unique in a sense that they are 
tightly connected to the QCD-vacuum fluctuations  with $0^{+},0^{-}$
singlet  quantum numbers. 
Manifestation of the uniqueness  
can be seen, in particular, in the existence of the axial anomaly ($0^{-}$ 
channel) and the trace anomaly ($0^+$ channel).
 
Well-known example  where this uniqueness shows up
is a large magnitude of the  strange content of the nucleon. 
In formal terms one can show  that   the matrix
element $\la N|\bar{s}s|N\ra$
    has the same order of magnitude as 
valence matrix element $\la N|\bar{d}d|N\ra$.
  We shall give a QCD-based
explanation of why a naively expected
suppression is not present there. 
After that, using an intuition
gained from this analysis,
 we turn into our main subject:
non-valence matrix elements in heavy hadrons.
 
We should note from the very
beginning of this letter that the ideology and methods (unitarity,
dispersion relations, duality, low-energy theorems) we use  are motivated
by QCD sum rules. However we do not use the QCD sum rules in the common 
sense.
Instead, we reduce one complicated problem 
(the calculation of non-valence nucleon matrix elements)
to another one (the behavior of some vacuum correlation functions
at low momentum transfer).
One could think that such a reducing of one problem to another one
(may be even more complicated) does not improve  our  understanding
 of the phenomenon. However, this is not quite true:
The analysis of the vacuum correlation functions  with vacuum quantum 
numbers,
 certainly, is a very difficult problem. However 
some nonperturbative information based on the low energy theorems
is available for such a correlation function.
Besides that, one and the same vacuum correlation functions 
enters into the different physical characteristics.
So, we could extract the unknown correlation function, let say,
from $\la N|\bar{s}s|N\ra$ and use this information in evaluation of
the matrix element we are interested in:
$\la \Lambda_b|\bar{s}s|\Lambda_b\ra$.
   Such an approach  gives a chance to estimate some interesting quantities.

    \section{Strangeness in the nucleon.   }
 Let us start from the standard arguments 
(see e.g. the text book \cite{Donoghue1})
showing a large magnitude of
of   $\la N|\bar{s}s|N\ra$. Arguments are based on 
the results of the fit to the data on $\pi N$ scattering  
 and they lead to the following estimates for the so-called $\sigma$ term
\cite{Gasser}:
\be
\label{1}
\frac{m_u+m_d}{2}\la p|\bar{u}u+\bar{d}d|p\ra=(45 MeV).
\ee
(Here and in what follows we omit kinematical structure like
$\bar{p}p$ in expressions for matrix elements.).
 Taking the values of quark masses to be
$m_u=5.1\pm0.9MeV~,m_d=9.3\pm1.4MeV~,m_s=175\pm25MeV$ \cite{Leutwyler},
 from (\ref{1}) we have
\be
\label{2}
\la p|\bar{u}u+\bar{d}d|p\ra\simeq   6.2,
\ee
where we literally use the center points for   
all parameters in  the  numerical estimations. 
Further, assuming octet-type $SU(3)$ breaking 
to be responsible for the mass splitting in the baryon octet, we find 
\be
\label{5}
\la p|\bar{u}u|p\ra\simeq  3.5,~
\la p|\bar{d}d|p\ra\simeq  2.8,~
\la p|\bar{s}s|p\ra\simeq  1.4.
\ee
We should mention that the accuracy of these equations is not very
high. For example, the error in the value of the $\sigma$ term
already leads to an error of order of one in each matrix element
discussed above. Besides that, 
 chiral perturbation
  corrections also give noticeable 
contribution into matrix elements (\ref{5}), see \cite{Gasser}.
However, the analysis of possible errors 
in eq. (\ref{5}) is not the goal of this paper.
Rather, we wanted to demonstrate that  these very simple calculations
explicitly show that the strange matrix element
    is not  small. Recent lattice calculations \cite{Liu}
  also support the large magnitude for the strange matrix element.

We would like to  interpret the relations (\ref{5})
  as a combination
of two very different (in sense of their origin)    contributions to the 
nucleon matrix element:
\be
\label{8}
\la p|\bar{q}q |p\ra\equiv \la p|\bar{q}q |p\ra_0 +\la p|\bar{q}q |p\ra_1,
\ee
where index $0$ labels a (sea) vacuum contribution and index $1$ a valence
contribution  for a quark $q$. In what follows we assume 
that the vacuum contribution which is 
related to the sea quarks is the same for all light quarks $u,d,s$.
Thus, the nonzero magnitude for the strange matrix elements
comes exclusively from the vacuum fluctuations. At the same time,
the matrix elements related to the valence contributions are equal to
\be
\label{9}
\la p|\bar{u}u |p\ra_1\simeq  2.1,~~
\la p|\bar{d}d |p\ra_1\simeq 1.4.
\ee
These values are in remarkable agreement with the numbers $2$ and $1$, which
one could expect from the naive picture of non-relativistic 
constituent quark model. 
 In spite of the very rough estimations presented above,
we believe we   convinced a reader that :
{\bf a)} a magnitude of the nucleon matrix element 
for $\bar{s}s$ is not small;
{\bf b)} the large value  for this matrix element is due to the
nontrivial QCD vacuum structure where vacuum expectation values
of $u,d,s$ quarks are developed and they have   the same 
order in magnitude:
$\la 0|\bar{d}d |0\ra\sim \la 0|\bar{u}u |0\ra
\sim \la 0|\bar{s}s |0\ra$.

Once we realized that the phenomenon under discussion
is related to the nontrivial 
vacuum structure, it is clear that the best way to 
understand such a phenomenon is  to use some method where QCD vacuum 
fluctuations and hadronic properties are strongly interrelated.
We believe, that the most powerful analytical nonperturbative method
which exhibits these features is 
the QCD sum rules approach \cite{Shif1},\cite{Shif2}.
\subsection{Strangeness in the nucleon and  
QCD vacuum structure.}
To calculate $\la N|\bar{s}s|N\ra$ using the QCD -sum rules approach, we
consider the following vacuum correlation function \cite{Khriplovich}:
\be
\label{11}
T(q^2)=\int e^{iqx}dxdy\la 0|T\{\eta(x),\bar{s}s(y), \bar{\eta (0)}\}|0\ra
\ee
at $-q^2\rightarrow\infty$. Here $\eta$ is an arbitrary current
with nucleon quantum numbers. In particular, this current may be chosen in the standard
form $\eta=\epsilon^{abc}\gamma_{\mu}d^a(u^bC\gamma_{\mu}u^c)$.  
For  the future 
convenience we consider the unit matrix kinematical structure in (\ref{11}).

 Let us note that  due to the absence of the $s$ -quark field in the nucleon
current $\eta$, any substantial contribution to $T(q^2)$ is 
connected only with non-perturbative, so-called induced vacuum condensates.
  Such a contribution arises from the region, when
some distances are large.
Thus, this contribution  can not be directly calculated in perturbative theory,
but rather should be  coded (parameterized)
in terms  of a bilocal operator $K$\cite{Khriplovich}:
\be
\label{14}
\la p|\bar{s}s|p\ra\simeq \frac{-m}{\la 0|\bar{u}u |0\ra }K ,
\ee
\be
\label{15}
K=i\int dy\la 0|T\{\bar{s}s(y), \bar{u}u(0)\}|0\ra   
\ee
where $m$ is the nucleon mass.
  For the different applications of this approach
where the bilocal operators play an essential role, see  
refs.\cite{Ioffe1},\cite{Balitsky},\cite{Ioffe}.

The main assumptions which   have been made in 
the derivation of this relation are the 
following. First, we made the standard assumption
about local duality for the nucleon. The second assumption is that
  the typical scales (or what is the same, duality intervals)
in the limit $-q^2\rightarrow\infty$ in the  
three -point    sum rules  (\ref{11}) and corresponding
two-point sum rules
\be
\label{13}
P(q^2)=\int e^{iqx}dx \la 0|T\{\eta(x), \bar{\eta (0)}\}|0\ra ,
\ee
 are not much different in magnitude from each other. 
In different words we assumed that
a nucleon saturates both correlation functions
with approximately equal duality intervals    $\sim S_0$. 
In this case the dependence on residues $\la 0|\eta|N\ra$ is canceled out
in the ratio of those correlation functions
and we are left with the matrix element $\la p|\bar{s}s|p\ra$ (\ref{14})
we are interested in.

One can estimate the value of $K$ by expressing this in terms of some
vacuum condensates \cite{Khriplovich}:
\be
\label{17}
K\simeq\frac{18}{b}\frac{\la\bar{q}q\ra^2}
{\la \frac{\alpha_s}{\pi}G_{\mu\nu}^2\ra}\simeq 0.04 GeV^2
\ee
where
$b=\frac{11}{3}N_c-\frac{2}{3}N_f=9 $ and we use the standard
values for the   condensates \cite{Shif1}, \cite{Shif2}:
$$\la \frac{\alpha_s}{\pi}G_{\mu\nu}^2\ra\simeq 1.2 \cdot 10^{-2}GeV^4~~~
\la\bar{q}q\ra\simeq -(250MeV)^3 .$$

The estimation (\ref{17} might be too naive, 
however, if  we  literally adopt  this estimate   for $K$,  
formula (\ref{14}) gives
the following expression for the 
nucleon expectation value for $\bar{s}s$
\be
\label{18}
 \la p| \bar{s}s  |p\ra \simeq
-m\cdot\frac{18}{b}\frac{\la\bar{q}q\ra }{\la \frac{\alpha_s}{\pi}G_{\mu\nu}^2\ra}
\simeq 2.4  ,
\ee
which is not far away from   ''experimental result" (\ref{5}).
Having in mind a large  uncertainties in those equations,
  we interpret 
an approach which leads to the final formula  (\ref{18})
as a very reasonable method 
for estimation of non-valence  matrix elements.

It is very important  that  our following  
formulas for the non-valence content
in   heavy quark system ( next section  ) will be expressed
in terms of the {\bf same} correlator $K$.
Therefore,
 we could   use   formula (\ref{14}) in order 
to extract the corresponding value
 for $K $ from experimental data
instead of using our estimation (\ref{17}). In this case
$K$ is given by
\be
\label{k}
K\simeq-\frac{1}{m}\la p|\bar{s}s|p\ra\la 0|\bar{u}u |0\ra 
\sim 0.025GeV^2 .
\ee
 Let us stress: we are not pretending to have made
a reliable calculation of the matrix element $ \la p| \bar{s}s  |p\ra$
here. Rather, we wanted to emphasize on the    qualitative picture which 
demonstrates the close relation 
between {\it non-valence matrix elements and QCD vacuum structure}. 
This is the lesson
number one. More lessons to be learned will follow.
\section{Zweig rule violation  in the vacuum channels. Lessons.}
The result (\ref{5},\ref{18})
means that   
 $s$ quark  contribution into the nucleon mass is not small. Indeed,
by definition
\be
\label{22}
m=\la N|\sum_{q}m_q\bar{q}q|N\ra-
\frac{b}{8}\la N| \frac{\alpha_s}{\pi}G_{\mu\nu}^2 |N\ra,
\ee
where  sum   is  over all light quarks $u, d, s$.
Adopting the values  
for $\la p| \bar{s}s  |p\ra \simeq   1.4$ and $m_s\simeq 175 MeV$ \cite{Leutwyler},
one can conclude that a noticeable  part of the nucleon mass
  (about $200\div 300 MeV$) is due to the strange quark. We have 
mentioned this, well known result,  
in order to emphasize
 that the same phenomenon
  takes place 
(as we argue in the next section)
in heavy quark system. Namely, we shall see that  $s$ quark contribution 
 to   $\bar{\Lambda}\equiv m_{H_Q}-m_Q|_{m_Q\rightarrow\infty}$ for
heavy hadron $H_Q$  
  is not small. 
This result is in a variance with the standard Zweig rule expectation
  predicting that  any non-valence matrix element is suppresed
in comparison with a similar in structure, but valence one.

The method presented above gives a very simple 
 QCD-based physical explanation of why  the Zweig rule in
  the scalar and pseudoscalar
channels  is badly broken  and at the same time, in the vector channel the 
Zweig rule works well. In fact, we reformulated the original
problem of the calculating of a non-valence matrix element
in terms of some {\it vacuum} nondiagonal correlation function
$ \sim \la 0|T\{\bar{s}\Gamma s(x), \bar{u}\Gamma u(0)\}|0\ra$
with a Lorenz structure $\Gamma$. 
 
In particular, the matrix element 
$\la N|\bar{s}\gamma_{\mu}s|N\ra$ 
is reduced to the analyses 
of the nondiagonal correlation function 
$\int dx \la 0|T\{\bar{s}\gamma_{\mu}s(x), \bar{u}\gamma_{\nu}u(0)\}|0\ra$,
which is expected to be very small
in comparison with the diagonal one
$\int dx \la 0|T\{\bar{u}\gamma_{\mu}u(x), \bar{u}\gamma_{\nu}u(0)\}|0\ra$.
Therefore, the corresponding matrix element 
 as well as
  the   coupling constant $g_{\phi NN}$ are also small. 
  In terms of   QCD
such a smallness corresponds to the numerical suppression   ( of order $10^{-2}-10^{-3}$)of the nondiagonal correlation function in comparison
with the diagonal one,
see QCD-estimation in \cite{Shif1}.

In the scalar and pseudoscalar channels the diagonal and non-diagonal 
correlators 
have the same order of magnitude; therefore, no suppression occurs.
This is the cornerstone 
of the paper and is the fundamental explanation of the phenomenon we are discussing
here. Specifically, magnitude of correlator $K$ is not 
changing much if we replace $s$ quark to $u$ quark in formula (\ref{15}).

 Of course, it is in contradiction
with large $N_c$ (number of colors) 
counting rule where a non-diagonal correlator
should be suppressed. The fact    that the naive counting
of powers of $N_c$ fails in channels with total spin $0$ is well-known:
quantities small in the limit $N_c\rightarrow\infty$ turn out to be large
and vice versa. This is manifestation of the phenomenon discovered
in ref.\cite{Novikov}: not all hadrons in the real world are equal
to each other. 

 Because the issue of the violation of $N_c$ counting rule (Zweig rule)
is so important
and because   all our results are based on this violation, we belive
it is appropriate to give more examples
(explanations)  where naive $N_c$ counting fails.
We hope that arguments presented below convince a reader that effect
we are talking about is not extraordinary one, but rather
is a very common phenomenon if we have  dealt with   $0^{\pm}$ vacuum channels.

We follow \cite{Novikov} and introduce the following ratio  
\be
\label{r}
r=\frac{\la 0|\frac{b}{8} \frac{\alpha_s}{\pi}G_{\mu\nu}^2 |2 gluons\ra^2}
{\la 0|\frac{b}{8} \frac{\alpha_s}{\pi}G_{\mu\nu}^2 |\pi\pi+KK+\eta\eta\ra^2}.
\ee
This ratio is very convenient since all
normalization factors due to phase volume cancel out.
Besides that, it is not difficult to find that this ratio is proportional
to $N_c^2$ because of the suppression of an amplitude creating a
pair of mesons in comparison with a creation of   gluon pair.
This ratio can be explicitly calculated from low energy theorem
and is given by \cite{Novikov}:
\be
\label{r1}
r=\frac{N_c^2-1}{16\ln^2(\frac{M}{\Lambda_{QCD}})},
\ee
where $ M\sim 1GeV $ is invariant mass of the pair of mesons, small enough 
for low energy expansion to be valid, large enough to have a small
coupling constant $\alpha_s(M)<1$.
In accordance with general rules $r\sim N_c^2\rightarrow\infty$ in the large
$N_c$ limit. However, for the real world with $N_c=3$ the ratio is small rather
than large, $r\sim0.1\ll 1$.
This is a clear indication  of the anomalously strong coupling in $0^+$ 
vacuum channel. 

The same phenomenon is responsible
for large  $\eta'$mass. Indeed, as was noted by Witten \cite{Witten}, unlike
all normal mesons whose masses are $N_c$ independent, the $\eta'$ mass
vanishes in the large $N_c$ limit. However, in the real world
this rule is badly broken, where $\frac{m^2_{\eta'}}{m^2_{\rho}}\simeq 1.8$.
The reason is the same as before -- the anomalously strong coupling in $0^-$ 
 vacuum channel.

Our next argument is as follows. It has been known
for a while that in some special cases\footnote{The   guess that it happens
exactly in the vacuum  $0^{\pm}$ channels is correct.}
the QCD sum rules do not work. In particular,  in the 
pseudoscalar quark channel  the standard  QCD sum rule  
for the correlation function \footnote{Note, that
(\ref{pion})  is very similar to our correlator $K$(\ref{15}).}
\be
\label{pion}
T(q^2)=\int e^{iqx}dx\la 0|T\{J^{\dagger}(x),   J(0)\}|0\ra ,~~ 
J=\bar{u}i\gamma_5d
\ee
{\bf can not} reproduce the residue $\la 0|J|\pi\ra\simeq f_{\pi}2GeV$ 
which is known exactly from
PCAC, and which has much bigger scale than QCD sum rule can provide.
So called direct instanton contributions play
a decisive role in these channels \cite{Novikov}. 

Another manifestation of the same phenomenon is 
somewhat different behavior of the four-quark condensates  
$\la \bar{q}\Gamma q\bar{q}\Gamma q\ra$ 
with   different Lorenz and color structure ($\Gamma$ in this formula
denotes a combination any of the $\gamma$ and   $\frac{\lambda^a}{2} $ or $\bf{1}$
matrices from the color and flavour groups).
 The so-called factorization hypothesis (which is justified 
in the large $N_c$ limit )works perfectly well for the vector
and axial-vector cases ($\Gamma=\gamma_{\mu},~\gamma_{\mu}\gamma_5$)\cite{Chetyrkin},
but does not work in general, see Shifman's comment
paper on this subject in the book\cite{Shif2}. 
In particular, the vacuum condensate
$\la \bar{u}\sigma_{\mu\nu}\lambda^a u\bar{d}\sigma_{\mu\nu}\lambda^ad\ra$
is not small
 in spite of the fact that the factorized value is exactly zero
\footnote{Actually, it has the same order of magnitude as 
the condensate 
$\la \bar{d}\sigma_{\mu\nu}\lambda^au \bar{u}\sigma_{\mu\nu}\lambda^a d\ra$
and both condensates are much bigger than the factorization hypothesis
predicts\cite{Zhitnitsky}.}. Such a behavior of condensates has   a qualitative 
explanation  based on 
the properties of the fermion zero modes
within the instanton approach, see e.g.review \cite{Shuryak1}.
 
The same instanton picture gives also a qualitative explanation
of the  enhancements mentioned above in the $0^{\pm}$ channels. 
  In fact, the forementioned Zweig rule violation is related to 
the fermion  zero modes, which always accompany an instanton.
 As is known, those zero modes are very selective in a sense
of the quantum numbers they carry on: they do contribute, let say, to the 
scalar correlator
  and they do not contribute
to the vector one. 

One could estimate the correlator $K$ (\ref{15}) using  the 
instanton ideas of ref.
\cite{Shuryak1}. The numerical result obtained in this 
way will be very close to our estimation (\ref{17}).
However, in spite of the very attractive  and  simple  
picture of the QCD vacuum structure advocated in ref.
\cite{Shuryak1}, 
it is very difficult to estimate an error 
of such a calculation. Therefore, we prefer to use 
a less direct, but more solid approach based on the
low-energy theorems. In this case    
the enhancement of the  vacuum channels (at least qualitatively) can be easily
understood.

Instead of the analysis of   the original correlator (\ref{15})
which enters into our formulae,
we introduce the following correlation function containing
a heavy quark $Q$:
\be
\label{Q}
K^{(Q)}=i\int dy\la 0|T\{\bar{Q}Q(y), \bar{u}u(0)\}|0\ra   .
\ee
 In the limit when a quark $Q$ is  very heavy, 
the correlator $K^{(Q)}$ can be calculated exactly!
Indeed, in this limit, one can use the standard 
operator product expansion $\sim (1/m_Q)^n$ in order to express the
quark operator $\bar{Q}Q $ in terms of the gluon operators\cite{Shif2}:
\be
\bar{Q}Q=-\frac{\alpha_s}{12m_Q\pi}G_{\mu\nu}^2  +
c\frac{G_{\mu\nu}^3}{m_Q^3}+...  .
\ee
 The correlation function $K^{(Q)}$ then takes the form
\be
\label{Q1}
K^{(Q)}= 
   -i \frac{1}{12m_Q}
\int dy\la 0|T \{\frac{\alpha_s}{\pi}G_{\mu\nu}^2(y) ,\bar{u}u(0)\}|0\ra 
+O(1/m_Q^3)
\ee
where we mainly interested in 
   the leading term    $\sim 1/m_Q$.
Fortunately, the obtained correlation function is known exactly   \cite{Novikov}:
 \be
\label{Q2}
i\int dy\la 0|T\{\frac{\alpha_s}{\pi}G_{\mu\nu}^2(y) ,\bar{u}u(0)\}|0\ra 
=\frac{8d}{b}\la\bar{u}u\ra ,
\ee
where $d=3$ is the dimension of the operator $\bar{u}u$ and $b=\frac{11N_c}{3}-\frac{2N_f}{3}=9$.
Finally, we get the following expression for the correlation function
we are interested in:
\be
\label{Q3}
K^{(Q)}=i\int dy\la 0|T\{\bar{Q}Q(y), \bar{u}u(0)\}|0\ra 
  =-\frac{2}{9}\frac{\la\bar{u}u\ra}{m_Q}+O(1/m_Q^3).
\ee
This is exact formula for large $m_Q$.

Few remarks are in order.
First, formula (\ref{Q3}) for $K^{(Q)}$ shows the correct sign $``+"$
which is expected for the light $s$-quark (\ref{k},\ref{17}).
This formula  also demonstrates the correct $N_c$ dependence at large $N_c$:
The correlator (\ref{Q3}) is of  order of one\footnote{
Let us remind that  $\la\bar{u}u\ra\sim N_c, ~~ b\sim N_c$. Therefore,
the combination on the right hand side of eq. (\ref{Q2}) is
of order $\frac{1}{b}\la\bar{u}u\ra \sim 1$.} 
rather than 
  $N_c$   expected for a diagonal quark correlation function. 
 One could estimate 
the next $1/m_Q^n,~ n>1$ corrections in the eq.(\ref{Q3}) with the result
that  the series   blows up when
$m_Q \leq 300\div 400 MeV$. Of course, this
result was expected from the very
beginning:  
one can not take the limit $m_Q\rightarrow m_s$ in the 
expansion like (\ref{Q3}). 
However, from the general consideration we expect that the correlator $K^{(Q)}$
is a monotonic function of $m_Q$ in the extended region
of $m_Q$ (except the region of the extremely small $m_Q\leq 30 MeV$ where the 
chiral perturbation theory predicts somewhat different behavior\cite{Novikov}).

The main goal 
of the present analysis of the correlator $K^{(Q)}$ is not a numerical
estimation (which is strongly model dependent magnitude in the
interesting region of $m_Q\simeq m_s\simeq 150 MeV$).
 Rather, we want to give a qualitative explanation of the enhancement
in the vacuum channels by analysing this correlator.
 On the qualitative level, one could expect   from the perturbative
analysis that the correlation function (\ref{Q})
should be suppressed by a factor of $\alpha_s^2$.
Indeed, an annihilation of the quark $Q$ into two gluons
and a creation of a pair with a different  flavour $u$ is suppressed in
perturbative calculation as $\alpha_s^2$.
Our   formula (\ref{Q3}) shows that this naive estimation
is wrong: no any suppression occurs in the exact formula.

It is clear why our intuition, based on the perturbative calculations is failed:
transition, we are talking about, is the large distance phenomenon.
Therefore, the perturbative analysis can not be applied to such an amplitude.
This statement
can be easily understood from the analysis of exact low-energy theorem (\ref{Q2}),
where a similar factor $\alpha_s^2$ has disappeared from the 
right hand side of the equation.

Interpretation of the disappearing  of this factor 
$\alpha_s^2$ is very simple: At large distances 
the most important configurations which 
are responsible for the transition like (\ref{Q2}) 
have an enhancement like $G_{\mu\nu}\sim\frac{1}{g}$.
Therefore,   semiclassical configurations
with  $G_{\mu\nu}\sim\frac{1}{g}$   
  saturate   the corresponding low-energy theorems;
they clearly can not be seen in perturbative analysis.
This remark closes our qualitative analysis of the 
correlation
function $K$ (\ref{15}). As we discussed earlier, one can not
use formula (\ref{Q3}) for the quantitative calculations for
$m_s\simeq 175MeV$. However, if we literally adopt 
this formula for $K$ with the assumption about its monotonic behavior
formulated above, we get
\be
\label{Q4}
K^{(Q)}=i\int dy\la 0|T\{\bar{Q}Q(y), \bar{u}u(0)\}|0\ra 
  \simeq -\frac{2}{9}\frac{\la\bar{u}u\ra}{m_Q}\rightarrow 
0.02 GeV^2 ~~\\
 at ~~~m_Q\simeq m_s\simeq 0.175 GeV, ~~~~~~~~~~~
~~~~~~~~~~~~~~~~~~~~~~~~~~~~~~~~~~~~\nonumber
\ee
which is very close to the ``experimental" value (\ref{k}).

Let us stress: we are not pretending to have made a reliable calculation
of the correlation function $K$ here. Rather, we wanted to emphasize
on the enhancement mechanism of the vacuum channels which could be understood
from the analysis of the low-energy theorems. This analysis also shows that
the corresponding enhancement is due to some semiclassical configurations
 in the functional integral with  $G_{\mu\nu}\sim\frac{1}{g}$. 
 
Finally, it is fair to say, that the limit of large $N_c$ nicely
explains  a lot of empirical regularities. The Zweig rule is particular example
of this kind. However, in vacuum channels this naive counting 
rule does not work. Therefore,
we should not be       surprised if we find
some strong deviation from the naive picture in  $0^{\pm}$ 
vacuum channels. Relatively large magnitude for the correlation
function $K$ (\ref{15}) (which is fundamentally important 
parameter for our estimations),
is another manifestation of the same kind.

\section{Heavy hadrons}
In this section we shall apply the ideas described above
for the calculation  of the non-valence matrix element
$\la \Lambda_b| \bar{s}s|\Lambda_b\ra$. It should be considered 
as an explicit
demonstration of the general idea (formulated in the introduction)
that a non-valence component
($\bar{s}s$) in a heavy quark system ($\Lambda_b\sim
bud$) could be large and comparable with valence matrix element
like $\la \Lambda_b| \bar{u}u|\Lambda_b\ra$. We notice, that
a similar conclusion was obtained previously
in the toy model of two-dimensional $QCD_2(N)$
\cite{Frishman}.

We start from the definition 
of the fundamental parameter $\bar{\Lambda}$\cite{Luke}
of HQET (heavy quark effective theory),
 see e.g. nice review paper \cite{Shifman2}:
\be
\label{23}
\bar{\Lambda}\equiv m_{H_Q}-m_Q|_{m_Q\rightarrow\infty}
\ee
All hadronic characteristics in HQET should be expressed in terms of 
$\bar{\Lambda}$ which is defined as the following matrix element: 
\be
\label{24}
\bar{\Lambda} =
\frac{1}{2m_{H_Q}}\la H_Q|\sum_{q}m_q\bar{q}q
+\frac{\beta(\alpha_s)}{4\alpha_s}G_{\mu\nu}^2|H_Q\ra  .
 \ee 
Numerically $\bar{\Lambda}\sim 500 MeV$ \cite{Braun1}.

Now we can use the same  technique (we have been using
in the previous section)
  to estimate the strange quark contribution into the mass of a heavy hadron:
\be 
\label{s}
\bar{\Lambda}(s)=\frac{1}{2m_{H_Q}}\la H_Q| m_s\bar{s}s|H_Q\ra.
\ee
Lessons we learned from the 
similar calculations
   teach us that this matrix element might be 
large enough.

 Technically, to calculate $\la \Lambda_b| \bar{s}s|\Lambda_b\ra$
 we use the same approach we described in Section 2; namely we
consider the following vacuum correlation function:
\be
\label{b1}
T(q^2)=\int e^{iqx}dxdy\la 0|T\{\eta(x),\bar{s}s(y), \bar{\eta (0)}\}|0\ra
\ee
  Here $\eta =\epsilon^{\alpha\beta\gamma}
(u^{T}_{\alpha}C\gamma_5b_{\beta})d_{\gamma}$ is the current
with $\Lambda_b$ quantum numbers\footnote{ To be more precise,
two baryons: $\Lambda_b (I=0)$ as well as $\Sigma_b(I=1)$ contribute
to this  correlation function. However, for qualitative analysis we assume 
that their matrix elements are similar. Therefore, in order to simplify
things,  we do not separate those states.}.
It is much more convenient in the case of heavy quark, 
to use
heavy quark expansion within QCD sum rules,
 as it was suggested for the first time 
in ref.\cite{Shuryak}. In this case, instead of external parameter
$q_{\mu}$, one should introduce parameter $E$ in the 
following way: $q_{\mu}=
(m_Q+E,~0~,~0~,~0~)$. Similarly, the resonance energy 
is defined as  $m_{H_Q}=m_Q+E_r$ etc. Therefore,
all low energy parameters do not depend on
$m_Q$  and they scale like $\bar{\Lambda}$ at large $m_Q$.

  Let us note that, similar to the nucleon case,
 due to the absence of the $s$ -quark field in the  
current $\eta$, the most important contribution to $T(q^2)$ comes 
from the induced vacuum condensates $K$.
We should consider, along with the analysis of the correlator  (\ref{b1}),  
the following   two-point correlation  function 
\be
\label{b2}
P(q^2)=\int e^{iqx}dx \la 0|T\{\eta(x), \bar{\eta (0)}\}|0\ra .
\ee
 As before,  
 we assume that
$\Lambda_b$ baryon saturates both correlation functions (\ref{b1}, \ref{b2})
with approximately equal duality intervals    $\sim S_0$. 
In this case the dependence on residues $\la 0|\eta|\Lambda_b\ra$ is canceled out
in the ratio of those correlation functions
and we are left with the matrix element $\la \Lambda_b|\bar{s}s|\Lambda_b\ra$  
we are interested in.

This is the standard first step of any calculation
of such a  kind: Instead of direct calculation of a  matrix element,
we reduce the problem to the computation of some correlation function.
As the next step, we use  the duality and dispersion relations to relate a 
physical matrix element to the QCD- based formula for the corresponding 
correlation function. This is essentially the basic idea of the QCD sum rules.  
 
With this remark in mind, the calculations very similar to 
(\ref{11},\ref{14}, \ref{15}) bring us to the following
formula\footnote{Similar to the nucleon case,
we use the local duality arguments (so-called, finite energy sum rules)
to estimate the matrix element (\ref{s}). Besides that, we
use the standard technical
 trick\cite{Braun} which suggests to
use the combination $(E-E_r)T(E)$
in   sum rules (\ref{b1}) rather than $T(E)$ itself. 
This trick allows  to
exponentially suppress an unknown contribution from  
the nondiagonal transitions
which include higher resonances.}
 \be
\label{25}
 \frac{1}{2m_{\Lambda_b}}\la \Lambda_b|\bar{s}s|\Lambda_b\ra\simeq 
( \frac{3}{4}S_0+E_r)\frac{K}{-\la\bar{q}q\ra}\simeq 1\div 2,  
\ee  
where  $S_0$ and $E_r$ are duality interval and binding energy
for the lowest state  with given
quantum numbers \footnote{We use 
$S_0\sim E_r\sim (0.5\div0.7)GeV$ 
and $K\simeq 0.025GeV^2$ (\ref{k}) for numerical estimations.}.
This formula is direct analog of the expression (\ref{14})
we derived previously for the nucleon. 
In the course of   calculation we have made the same assumptions 
  we made  before, see previous section. Therefore,  
we believe we have the same accuracy as before 
which we estimate on the level   of  $50\%$.  The only difference
with formula (\ref{14}) is a replacement
of nucleon mass $m\simeq 1 GeV$ by a combination of  two   
parameters  $S_0$ and $E_r$ which   have the same order of magnitude
as nucleon mass.
In a sense, those parameters 
  are trivial kinematical factors which always have 
a hadronic $\sim 1 GeV$ scale. 

There is a  non-trivial factor in our formula which is very important
  for us  and deserves an additional explanation.  
  The fact is:
   the  nonperturbative correlation function $K$ which enters 
into the expression (\ref{25}) is the {\bf same} correlator  
we have been using   for the calculation $\la p| \bar{s}s  |p\ra $
(\ref{14}).
This factor is not small as
naively one could expect. It sets the scale of the phenomenon. 

Moral: If we accept the large value for 
$\la p| \bar{s}s  |p\ra $   we should also accept the
large value for 
\be
\label{26}
 \frac{1}{2m_{H_Q}}\la H_Q|m_s\bar{s}s|H_Q \ra 
\sim (200\div 300 ) MeV ,
\ee
as a consequence of   absence of any suppression 
for nondiagonal  correlator $K$.

Let us repeat: we are not pretending to have made
a reliable calculation of the matrix element $ \frac{1}{2m_{H_Q}}\la H_Q|m_s\bar{s}s|H_Q \ra $
here. Rather, we wanted to emphasize on the    qualitative picture which 
demonstrates a close relation 
between the matrix element (\ref{26}) and corresponding nucleon
matrix element (\ref{14}). Both those matrix elements
are related to each other and relatively large because
of the strong   fluctuations in vacuum $0^{\pm}$ channels.
  We can not calculate the nontrivial part (correlator $K$)
from the first principles. However, the analysis of different low-energy theorems
supports our expectation that its magnitude is large.
 For numerical estimations, we can extract a relevant information
from one problem in order to use this info somewhere else.
\section{ Conclusion}
We have argued that matrix element (\ref{26}) could be numerically large.
The arguments are  very similar to the case of strange   matrix element
over nucleon and based on the fundamental property of nonperturbative QCD
that there is no suppression for flavor changing amplitudes in the vacuum channels
$0^{\pm}$ (the Zweig rule in these channels is badly broken). Few 
consequences of the result (\ref{26}) are in order:

1. The value of $\bar{\Lambda}$ continues to be controversial, because the
QCD sum rules indicate that $\bar{\Lambda}\sim 0.5 GeV$ which 
  does not contradict to the lower
bound stemming from Voloshin's sum rules 
\cite{Voloshin},\cite{Shifman2}. At the same time 
 the lattice calculations give much smaller number:
$\bar{\Lambda}\sim 0.2-0.3 GeV$, see \cite{Shifman2} for more details.
The possible interpretation is: lattice definition of $\bar{\Lambda}$
does not correspond to the continuum theory because the
  $s$-quark contribution (\ref{26}) was not accounted properly.
It would be very interesting to calculate matrix element
(\ref{26}) in somewhat independent way; for example, 
in chiral perturbation theory or on the lattice (similar
to the nucleon calculation of ref.\cite{Liu}).

2. Scalar and pseudoscalar light mesons ($\eta$, $f_0$...)
 strongly interact with $\Lambda_b$; $\phi$ meson does not interact
with $\Lambda_b$.

3. We expect a similar situation for all heavy hadrons. Therefore,
for inclusive production of strangeless heavy hadrons  we expect   
some excess of strangeness in comparison with naive calculation.
However, we do not know how to estimate this effect in appropriate way.

We conclude with few general remarks:

4. A variation of the strange quark mass
 may  considerably change some
vacuum and hadronic characteristics. Therefore, 
the standard  lattice calculations of those characteristics using a
 quenched approximation     
is questionable  
simply  because such a calculation
 clearly not  accounting
the fluctuations of the strange (non-valence)quark. 
At the same time, the QCD sum rules approach clearly includes those
contributions implicitly. Indeed,
 all relevant  vacuum condensates (like $\la\bar{u}u\ra$)
which appear in the QCD sum rules approach for the 
non-strange hadrons   do depend on $s$ quark.

In fact, the correlator $K$  enters
to expression (\ref{14}), as well as it determines the variation
of the condensate $\la\bar{u}u \ra$ with $s$ quark mass:
\be
\label{19}
\frac{d}{dm_s}\la\bar{u}u \ra=
-i\int dy\la 0|T\{\bar{s}s(y), \bar{u}u \}|0\ra =-K\simeq -0.025GeV^2.
\ee
To understand how large this number is and in order to make 
some rough estimations, we assume that this behavior can be 
extrapolated from physical value $m_s\simeq 175 MeV$ till $m_s=0$.
In this case we estimate that
\be
\label{20}
\mid\frac{ \la\bar{u}u \ra_{m_s=175}-\la\bar{u}u \ra_{m_s=0}}
{\la\bar{u}u \ra_{m_s=175}}\mid\simeq 0.3.
\ee
Such a decrease of $\mid\la\bar{u}u \ra\mid$ by a $30\%$
as $m_s$ varies from $m_s\simeq 175 MeV$ to $m_s=0$ is a very important
consequence of QCD. Therefore, QCD sum rules approach implicitly
accounts an existence of strange quark in the theory.

5. An  analysis of the   low-energy theorems (similar to (\ref{Q3}))
might be useful tool for the future investigations on the Zweig rule violations
in different channels.

 \end{document}